\documentclass[%
 reprint,
 amsmath,amssymb,
 aps,
prc,
]{revtex4-2}

\usepackage{graphicx}
\usepackage{dcolumn}
\usepackage{bm}
\usepackage{hyperref}
\usepackage[mathlines]{lineno}

\begin{document}

\preprint{APS/123-QED}

\title{Contribution to differential $\pi^0$ and $\gamma_\mathrm{dir}$ modification\\ in small systems from color fluctuation effects}

\author{Dennis V. Perepelitsa}
\email{dvp@colorado.edu}
\affiliation{%
 University of Colorado Boulder \\
 Boulder, CO 80309
}%

\begin{abstract}
A major complication in the search for jet quenching in proton-- or deuteron--nucleus collision systems is the presence of physical effects which influence the experimental determination of collision centrality in the presence of a hard process. For example, in the proton color fluctuation picture, protons with a large Bjorken-$x$ ($x \gtrsim 0.1$) parton interact more weakly with the nucleons in the nucleus, leading to a smaller (larger) than expected yield in large (small) activity events. \\
\indent A recent measurement by PHENIX compared the yield of neutral pion and direct photon production in $d$+Au collisions, under the argument that the photon yields correct for such biases, and the difference between the two species is thus attributable to final-state effects (i.e., jet quenching). The main finding suggests a significant degree of jet quenching for hard processes in small systems. \\
\indent In this paper, I argue that the particular photon and pion events selected by PHENIX arise from proton configurations with significantly different Bjorken-$x$ distributions, and thus are subject to different magnitudes of modification in the color fluctuation model. Using the results of a previous global analysis of RHIC and LHC data, I show that potentially all of the pion-to-photon difference in PHENIX data can be described by a proton color fluctuation picture at a quantitative level before any additional physics from final-state effects is required. This finding reconciles the interpretation of the PHENIX measurement with others at RHIC and LHC, which have found no observable evidence for jet quenching in small systems.
\end{abstract}

\maketitle

\section{Introduction}

Measurements of hard processes in relativistic proton- or deuteron-nucleus ($p/d$+A) collisions at the Relativistic Heavy Ion Collider (RHIC) and the Large Hadron Collider (LHC) have a number of scientific purposes~\cite{Salgado:2011wc,Salgado:2016jws}, including a precision determination of the parton densities in nuclei (see Ref.~\cite{Eskola:2021nhw} for a recent example), constraints on dynamical processes in the initial state of the cold nucleus (such as those observed at lower energies~\cite{CLAS:2022asf}), and studies of the general interplay between soft and hard processes in collisions involving nuclei. Another key interest, given the robust experimental signatures of Quark Gluon Plasma-like behavior in these ``small'' systems~\cite{Nagle:2018nvi,Weller:2017tsr,ALICE:2016fzo,CMS:2018loe,ATLAS:2019vcm}, is the search for evidence of final-state interactions between the hard-scattered parton and the dilute system. If the parton shower is modified in the final state as it propagates through the produced system, it may lead to a number of effects, the most direct of which is the decreased production of jets or hadrons at fixed transverse momentum ($p_\mathrm{T}$), i.e., jet quenching~\cite{Tywoniuk:2014hta,Huss:2020dwe,Perepelitsa:2020pcf}. 

By analogy to large, nucleus-nucleus collision systems, one wants to search in the most ``central'', or highest-activity, collisions where the produced system is largest and longest-lived. Unfortunately, the experimental selection on centrality in small systems is challenging due to strong auto-correlation biases and non-trivial physics effects, with magnitudes larger than that required for a precision constraint on jet quenching effects. For events with a generic hard process, the selection is sensitive to upward multiplicity fluctuations (multiplicity vetoes), leading to an apparently enhanced yield in central (decreased yield in peripheral) collisions compared to that expected from the estimated number of nucleon--nucleon collisions, $\left<N_\mathrm{coll}\right>$, in the events~\cite{PHENIX:2013jxf,ALICE:2014xsp,Perepelitsa:2014yta,Loizides:2017sqq}. 

A separate phenomenon observed in data is that, in extreme kinematic regions characterized by Bjorken-$x$ $\gtrsim 0.1$, the pattern appears to reverse, with a significant depletion (enhancement) in the high-activity central (low-activity peripheral) events. This feature has been observed in measurements by ATLAS~\cite{ATLAS:2014cpa}, CMS~\cite{CMS:2014qvs}, STAR~\cite{STAR:2024nwm}, and PHENIX~\cite{PHENIX:2015fgy} in $p/d$+A events. In a recent measurement by ATLAS~\cite{ATLAS:2023zfx}, the produced dijet pair was used to estimate the parton-level kinematics in each event, strongly suggesting that the modifications follow a universal pattern in the Bjorken-$x$ of the proton, with their magnitude systematically increasing with $x$. A particular quantitative interpretation of these observations in data is the QCD color fluctuation model implemented in Refs.~\cite{Alvioli:2014eda,Alvioli:2017wou}, with references therein describing the theoretical development of this idea. The model proposes that proton configurations with a large-$x$ parton interact more weakly with nucleons in the nucleus than configuration-averaged protons, thus leading to a decreased centrality signal and the particular pattern of $x$-dependent modifications described above.

To get around these challenges, experimental searches of jet quenching in $p/d$+A collisions have therefore been performed in centrality-averaged (minimum-bias) collisions~\cite{CMS:2016xef,ALICE:2021est,ATLAS:2022kqu}, or using techniques such as jet--hadron correlations~\cite{ALICE:2017svf,ATLAS:2022iyq} and di-jet asymmetries~\cite{CMS:2014qvs,STAR:2024nwm} in central events, all giving null results for a detectable jet quenching-like signature. A recent PHENIX measurement~\cite{PHENIX:2023dxl} in 200 GeV $d$+Au collisions used a different technique and reported the relative yields of neutral pions ($\pi^0$'s) to those of direct photons ($\gamma_\mathrm{dir}$'s), under the argument that both are subject to the same centrality bias effects, and thus any difference in the degree of modification should be attributed to centrality-dependent final-state effects (i.e., which would not be felt by the photons). The PHENIX measurement reports the quantity $R_\mathrm{dAu,EXP}$, defined as the double ratio of the nuclear modification factors for $\pi^0$'s and $\gamma_\mathrm{dir}$, $R_\mathrm{dAu,\pi^0} / R_\mathrm{dAu,\gamma_\mathrm{dir}}$, i.e., conceptually defining $R_\mathrm{dAu,\gamma_\mathrm{dir}}$ to be unity for all centrality selections. The measurement reports $R_\mathrm{dAu,EXP}$ in the most central 0--5\% of $d$+Au events as $0.77 \pm 0.03 (\mathrm{stat}) \pm 0.13 (\mathrm{syst})$, where the latter is dominated by a global uncertainty due to the normalization of the $\gamma_\mathrm{dir}$ measurement. If interpreted as a jet quenching effect, the data are challenging to understand given the extensive null results described above. For example, the PHENIX data would require a  $\pi^0$ $p_\mathrm{T}$ spectrum shift of $\delta p_\mathrm{T} \approx -0.21$~GeV in the most central events at RHIC~\cite{AxelTalk}, corresponding to a fractional energy loss of $\delta p_\mathrm{T}/p_\mathrm{T} \approx 1$--$2\%$ in $d$+Au events, whereas the measurement by ATLAS in Ref.~\cite{ATLAS:2022iyq} sets an exclusion limit at 90\% confidence level of $\delta p_\mathrm{T}/p_\mathrm{T} < 1.4$\% for charged hadrons in $p$+Pb collisions at the LHC. However, these two measurements probe partons with different initial energies and collisions with different energies and geometries, and thus may not be in explicit tension. 

Notably, the $\pi^0$ and $\gamma_\mathrm{dir}$ were measured in the same $p_\mathrm{T}$ range of $7.5$--$18$~GeV which, since the $\pi^0$ carries only a fraction of the tree-level parton $p_\mathrm{T}$, actually corresponds to different Bjorken-$x$ ranges. Thus, the measurement technique defined by PHENIX is directly sensitive to the $x$-dependent color fluctuation effects described above, in addition to any jet quenching effects. In particular, the strength of the effect changes quickly with $x$ in the region of the measurement, and is thus only partially ``calibrated out'' with the $\gamma_\mathrm{dir}$ baseline. In this paper, I argue that the color fluctuation effect, as calculated using the results of Ref.~\cite{Alvioli:2017wou} with no additional model modifications or re-tuning to new data, can potentially explain the full centrality dependence of the observable defined by PHENIX. When accounting for it, the possible magnitude of any jet quenching effect is significantly smaller, reconciling the interpretation of the PHENIX measurement with the established constraints on jet quenching by the other RHIC and LHC experiments.

\section{Method}

The \textsc{Pythia} Monte Carlo event generator~\cite{Bierlich:2022pfr} was used to simulate the nucleon--nucleon sub-collisions in $d$+Au collisions which produce $\pi^0$'s and $\gamma_\mathrm{dir}$'s in the PHENIX kinematic selections~\cite{PHENIX:2023dxl}, and determine the distribution of Bjorken-$x$ values for these events. The specific selections are $p_\mathrm{T} = 7.5$--$18$~GeV and $\left|\eta\right| < 0.35$ for both species. \textsc{Pythia} version 8.307 was used to generate events at $\sqrt{s} = 200$~GeV, with a mixture of proton--proton, proton--neutron, and neutron--neutron collisions appropriate for $d$+Au. \textsc{Pythia} was configured with all HardQCD processes and a $\hat{p}_{\mathrm{T,min}}$ (minimum parton--parton transverse momentum exchange) of 5~GeV for the $\pi^0$ case. For the direct photon case, \textsc{Pythia} was configured with all PromptPhoton processes, also with $\hat{p}_{\mathrm{T,min}} = 5$~GeV, and photons which are the result of hadron decays were rejected (i.e., only those radiated by a quark were selected). For both samples, this $\hat{p}_{\mathrm{T,min}}$ threshold was checked that it does not introduce a kinematic bias in the final-state $p_\mathrm{T}$ range of interest. One of the beam hadrons was chosen to represent the proton or neutron in the deuteron of a $d$+Au collision and its Bjorken-$x$ value, defined in \textsc{Pythia} as the $x$ at which the PDF in the beam is defined~\cite{Bierlich:2022pfr}, was recorded.

\begin{figure}[!t]
\includegraphics[width=1.0\linewidth]{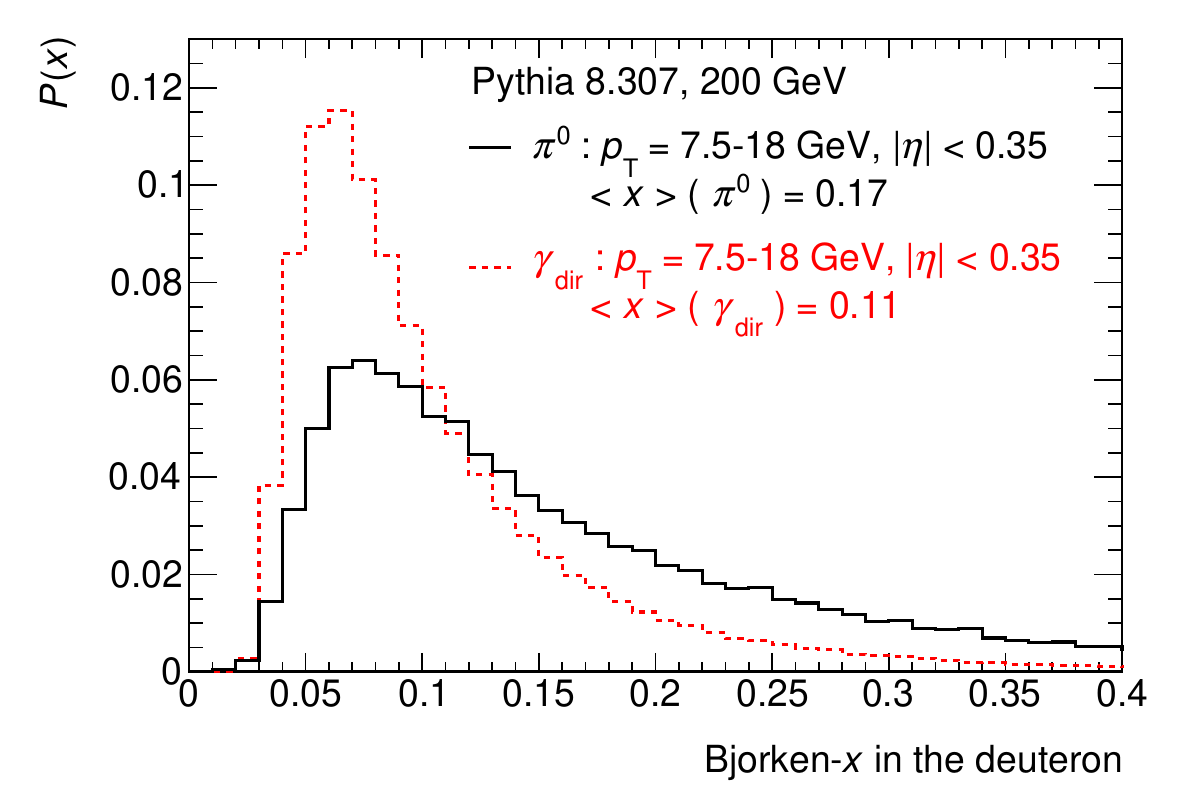}
\caption{\label{fig:x} Distribution in \textsc{Pythia}8 of the values of Bjorken-$x$ in the deuteron in 200 GeV $d$+Au collisions, for events producing $\pi^0$'s (solid black) or $\gamma_\mathrm{dir}$'s (dashed red) in the kinematic selection used by PHENIX in Ref.~\cite{PHENIX:2023dxl}.}
\end{figure}

Figure~\ref{fig:x} shows the distribution of Bjorken-$x$ values in the deuteron for events with a $\pi^0$ or $\gamma_\mathrm{dir}$ within the PHENIX kinematic selection. For  $\gamma_\mathrm{dir}$, the distribution is peaked near $x \sim 7.5 / 100$ = 0.075, where $7.5$~GeV and $100$~GeV are the minimum $\gamma_\mathrm{dir}$ $p_\mathrm{T}$ threshold and the beam energy, respectively, with a small tail of contributions from events with higher $x$. For $\pi^0$, the distribution has a significantly harder tail to large-$x$ values, reflecting the fact that they are produced by the fragmentation of hard scattered partons and thus carry only a fraction of their energy. Notably, the average $x$ value is more than 50\% larger for $\pi^0$- than $\gamma_\mathrm{dir}$-producing events, with 28\% (9\%) of the $\pi^0$'s ($\gamma_\mathrm{dir}$'s) produced in events with $x > 0.2$. 
In principle, the magnitude of the color fluctuation effect is expected to be different depending on whether the high-$x$ parton in the deuteron is a quark or a gluon. However, in these kinematics, the quark/gluon fraction is similar in the $\pi^0$ and $\gamma_\mathrm{dir}$ events and thus the difference in their $x$ distributions is the main distinction.

The distributions of $x$ values in Fig.~\ref{fig:x} were then folded with the results of the color fluctuation model in Ref.~\cite{Alvioli:2017wou} to produce a prediction for the results of the PHENIX measurement. First, the color fluctuation model paper primarily reports $R_\mathrm{CP}$, the ratio of nuclear modification in central to peripheral events, which must be converted to the $R_\mathrm{dAu,EXP}$ and other observables reported in the recent PHENIX paper. This is accomplished using the detailed information provided in an earlier PHENIX publication~\cite{PHENIX:2013jxf} on the geometric properties (e.g. the average number of colliding nucleon--nucleon pairs $\left<N_\mathrm{coll}\right>$) of centrality-selected $d$+Au events. Importantly, while the results here are presented as a function of the average $\left<N_\mathrm{coll}\right>$, the calculations in Ref.~\cite{Alvioli:2017wou} on which they are based were determined using the distribution of the $N_\mathrm{coll}$ values in each centrality bin. Second, the color fluctuation model presents results in exclusive Bjorken-$x$ bins, which are then combined in a weighted average with the $x$ distributions for $\pi^0$ and $\gamma_\mathrm{dir}$ events above. Since the model in Ref.~\cite{Alvioli:2017wou} did not consider the region $x \lesssim 0.1$, I treat the color fluctuation effect as negligible for events with $x$ in this range, i.e. for these all $R_\mathrm{dAu} = 1$. I highlight that the parameters of this model, determined using previous RHIC and LHC data, have not otherwise been adjusted or updated before the comparison to the PHENIX data of interest. 

\begin{figure}[!t]
\includegraphics[width=1.0\linewidth]{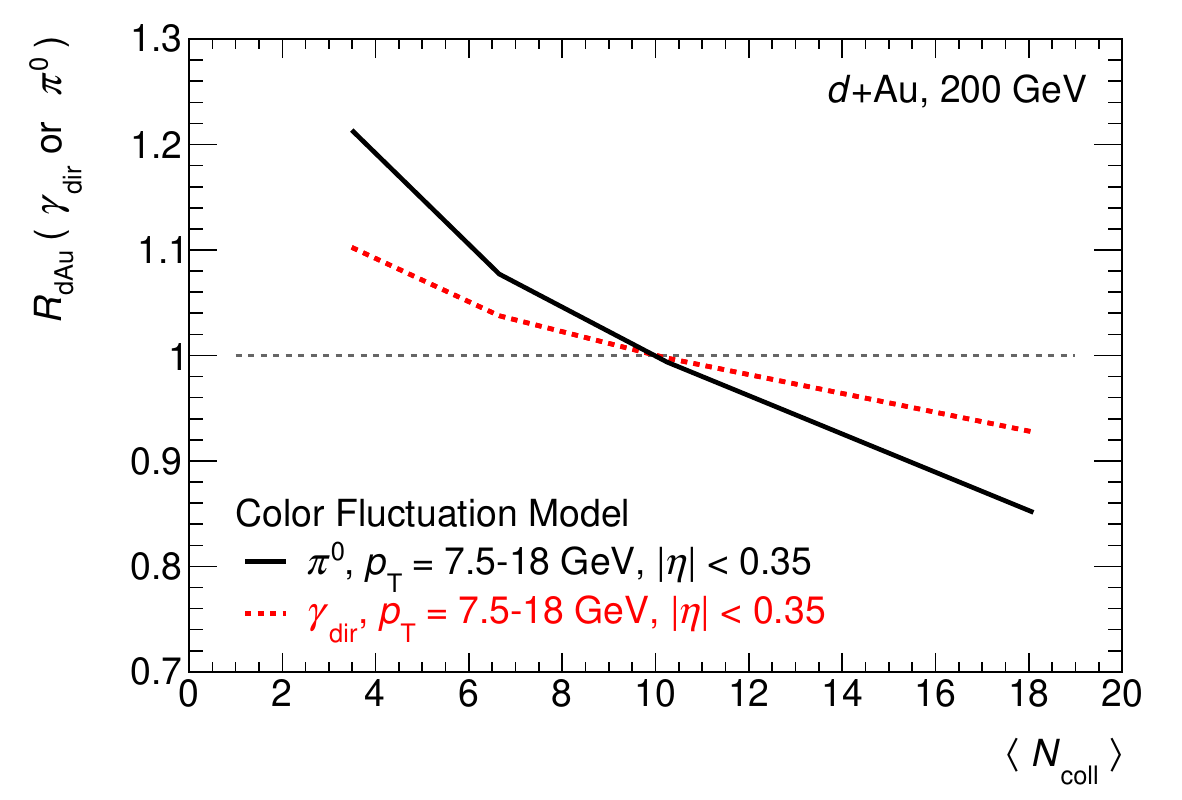}
\caption{\label{fig:model} Calculated nuclear modification factor $R_\mathrm{dAu}$ within the color fluctuation model as a function of the average number of colliding nucleon pairs, $\left<N_\mathrm{coll}\right>$. Calculations are shown for $\pi^0$'s (solid black) and $\gamma_\mathrm{dir}$'s (dashed red) matching the PHENIX kinematic selection. Since this model does not include any overall modification, the minimum-bias $R_\mathrm{dAu}$ is unity by construction.}
\end{figure}

Notably, the color fluctuation model in Ref.~\cite{Alvioli:2017wou} does not itself include any centrality-independent modification, i.e., $R_\mathrm{dAu}$ for minimum-bias events is unity by construction. However, there are expected to be small modifications of the minimum-bias rate for both species due to the nuclear parton densities~\cite{Eskola:2021nhw} and the ``isospin effect'' for direct photon production from the valence quark region~\cite{PHENIX:2012krx,ATLAS:2019ery}. Given the specific interest in the centrality dependence of the observable and the small magnitude of the minimum-bias-averaged effects compared to the large overall normalization uncertainties in PHENIX data, I do not model these effects explicitly in this work. If such effects were included, they would result in a global rescaling of all calculated $R_\mathrm{dAu}$ points in the same direction.

Fig.~\ref{fig:model} shows the calculated nuclear modification factors, $R_\mathrm{dAu}$, within the color fluctuation model, applied to the $\pi^0$ and $\gamma_\mathrm{dir}$ events in the PHENIX kinematics selection, and within the PHENIX centrality intervals. Since the minimum-bias $R_\mathrm{dAu}$ is unity by construction (see above), the modifications appear as a relative enhancement/suppression pattern around unity in peripheral/central events. Importantly, the magnitude of the relative modifications is stronger for $\pi^0$'s than for $\gamma_\mathrm{dir}$'s, reflecting the different $x$ distributions in Fig.~\ref{fig:x}. Indeed, in Ref.~\cite{Alvioli:2017wou}, the quantity $\lambda(x)$, which describes the $x$-dependent weakening of the inelastic nucleon--nucleon interaction strength, falls rapidly over the Bjorken-$x$ range relevant for $\gamma_\mathrm{dir}$ and $\pi^0$ production in the measurement, where it decreases from $\lambda \approx 0.8$ at $x \approx 0.1$ to $\lambda \approx 0.5$ at $x \approx 0.4$. 

\section{Comparison to Data}

\begin{figure}[!t]
\includegraphics[width=1.0\linewidth]{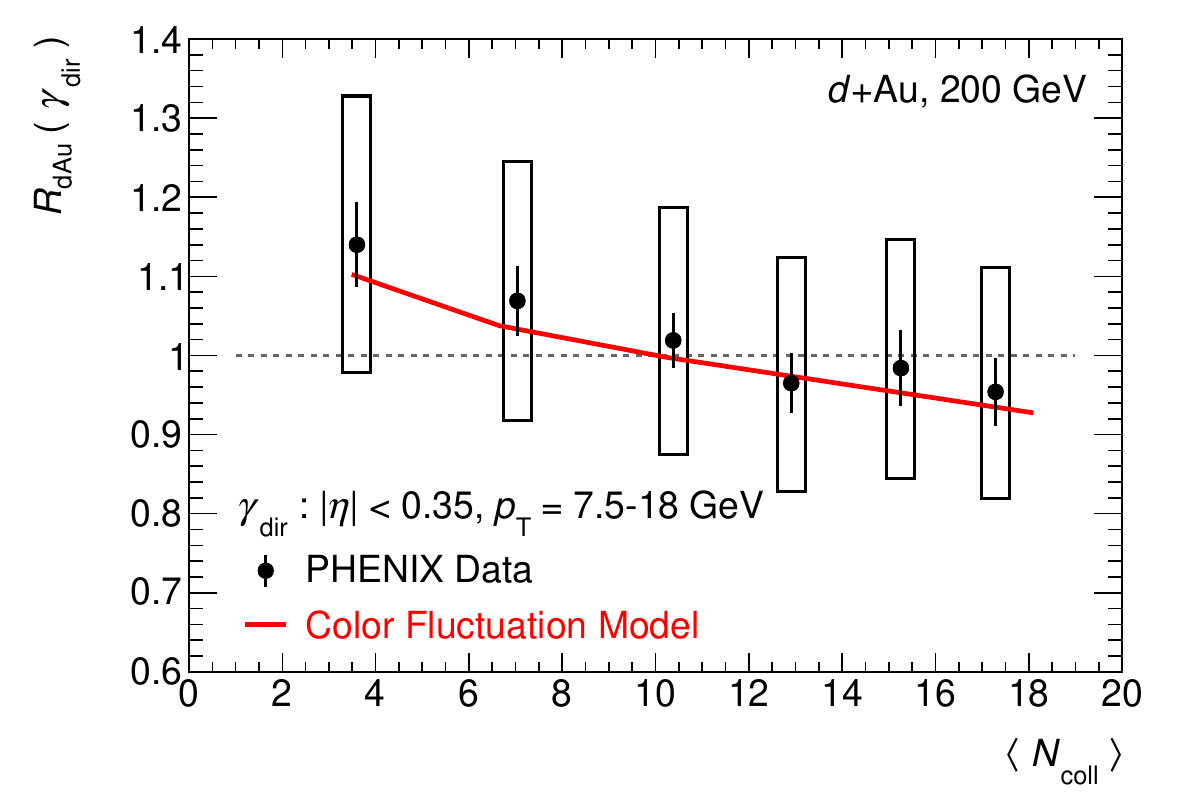}
\caption{\label{fig:data1} Comparison of the nuclear modification factor $R_\mathrm{dAu}$ for direct photons as a function of $\left<N_\mathrm{coll}\right>$, showing the measurement in PHENIX data (black points) and the calculation from the color fluctuation model (red line). The vertical bars and boxes around the data points indicate the statistical uncertainties and the overall 16.5\% normalization uncertainty, respectively.}
\end{figure}

Fig.~\ref{fig:data1} compares the calculated $R_\mathrm{dAu}$ for $\gamma_\mathrm{dir}$ integrated over the $p_\mathrm{T} = 7.5$--$18$~GeV range within the color fluctuation model to that measured by PHENIX, where the quantity is styled as  $(Y^{\gamma_\mathrm{dir}}_\mathrm{dAu}/Y^{\gamma_\mathrm{dir}}_\mathrm{pp})/N_\mathrm{coll}^\mathrm{Glauber}$. For this comparison, the measured minimum-bias $R_\mathrm{dAu}$ for $\gamma_\mathrm{dir}$ is essentially unity, and so no overall rescaling of the model calculation is applied. 
The model gives a good description of the modest centrality dependence of the data, with a relative enhancement and suppression pattern in going from peripheral to central events. As a reminder, the $\gamma_\mathrm{dir}$ selection typically results in events with $\left<x\right> \approx 0.11$, and thus the impact of the color fluctuation physics on this observable is modest in the $x$ range relevant for $\gamma_\mathrm{dir}$ production.

\begin{figure}[!t]
\includegraphics[width=1.0\linewidth]{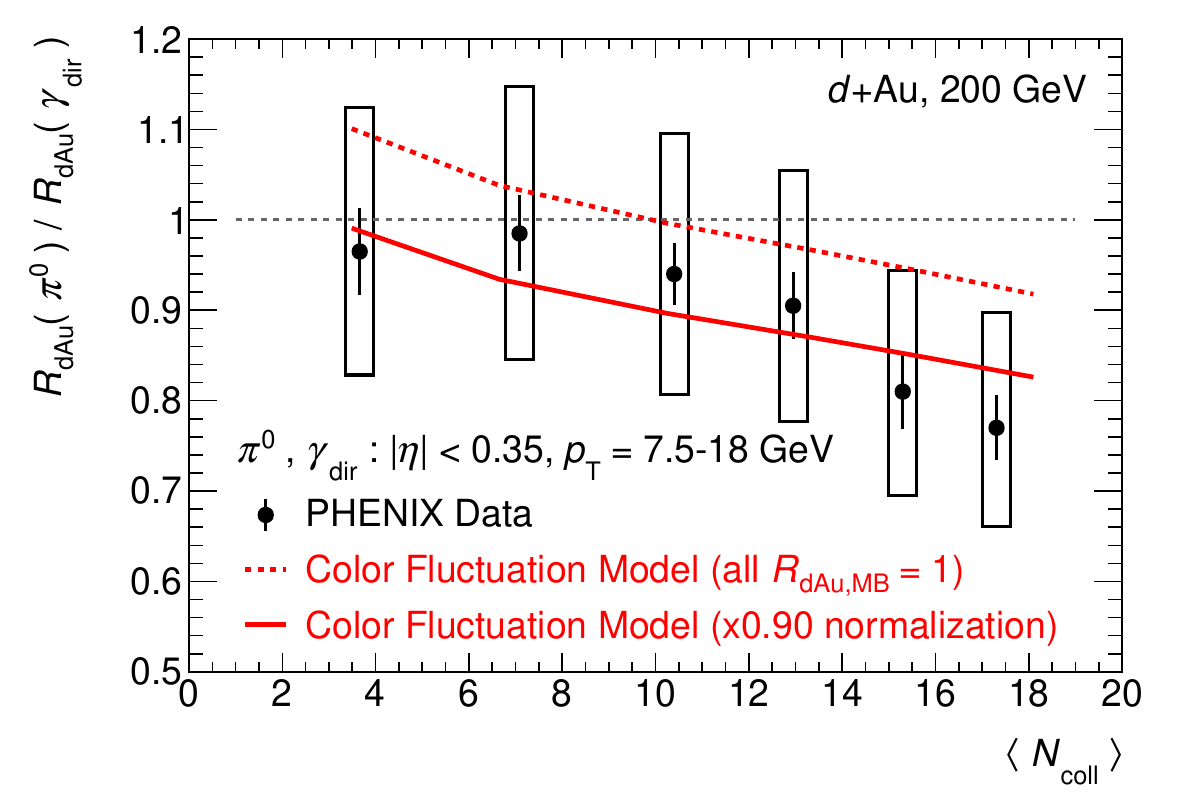}
\caption{\label{fig:data2} Comparison of the ratio of nuclear modification factors between that for neutral pions and direct photons, $R_\mathrm{dAu}(\pi^0)/ R_\mathrm{dAu}(\gamma_\mathrm{dir})$, as a function of $\left<N_\mathrm{coll}\right>$, showing the measurement in PHENIX data (black points) and the calculation from the color fluctuation model (red lines). The vertical bars and boxes around the data points indicate the statistical uncertainties and the overall 16.5\% normalization uncertainty, respectively. The dashed red line assumes that $R_\mathrm{dAu}$ in minimum-bias events for both $\pi^0$'s and $\gamma_\mathrm{dir}$'s is unity. The solid red line adjusts the calculation by a factor of 0.9, to better match the actually measured $R_\mathrm{dAu}(\pi^0)/ R_\mathrm{dAu}(\gamma_\mathrm{dir})$ in minimum-bias events by PHENIX.}
\end{figure}

Fig.~\ref{fig:data2} compares the measured and calculated ``double ratio'' of the $R_\mathrm{dAu}$ for $\pi^0$'s to that for $\gamma_\mathrm{dir}$'s, as a function of $\left<N_\mathrm{coll}\right>$, which is the main result of the PHENIX analysis. The dashed red line shows the result of the color fluctuation model, under the assumption that the $R_\mathrm{dAu}$ in minimum-bias events is unity for both species. The calculation gives a good, quantitative description of the data within its significant normalization uncertainty, including the slow decrease of the double ratio with increasing $\left<N_\mathrm{coll}\right>$. 

In the PHENIX analysis, the minimum bias $R_\mathrm{dAu}(\pi^0)/R_\mathrm{dAu}(\gamma_\mathrm{dir})$ is measured to be $0.92 \pm 0.02(\mathrm{stat})\pm0.15(\mathrm{syst})$. Ref.~\cite{PHENIX:2023dxl} notes that this value is consistent with unity within the large overall scale uncertainty, which is dominated by uncertainties in the $\pi^0$ and $\gamma_\mathrm{dir}$ $p$+$p$ references. Thus, the data-model comparison here has a significant degree of freedom in the overall normalization of the data. The best match between the model and the central values of the data is achieved by adjusting the model down by a factor $0.90$, shown as a solid red line in Fig.~\ref{fig:data2}. I note that this factor is close to the measured minimum-bias $R_\mathrm{dAu}(\pi^0)/R_\mathrm{dAu}(\gamma_\mathrm{dir})$ value of $0.92$, and is well within the normalization uncertainty of the data. This may alternatively be thought of as scaling the data up by approximately two thirds of its stated global uncertainty. With this normalization, the model now provides an excellent description of the central values of the PHENIX data.
While not explored here, I note that taking into account the uncertainties in the model parameters themselves~\cite{Alvioli:2017wou} may further improve the data-model agreement. Thus, no other physics effects, such as a significant centrality-dependent jet quenching, are needed to describe the PHENIX data.

In the future, several experimental avenues may help to separate the contribution of the color fluctuation effect from searches for the jet quenching physics of interest. First, one may design the kinematic selections with an aim to match the Bjorken-$x$ ranges accessed by the hadron/jet and photon processes, and thus cancel the impact of the color fluctuation effect. Second, as motivated in Ref.~\cite{McGlinchey:2016ssj}, the two effects are expected to have different sensitivities to changing the projectile, such as in $p$+Au or $^{3}$He+Au data previously recorded at RHIC. This has been performed for $\pi^0$'s in Ref.~\cite{PHENIX:2021dod}, but not yet for $\gamma_\mathrm{dir}$. Third, one could attempt a centrality selection based on measuring spectator neutrons in a zero-degree calorimeter (e.g. as used in Refs~\cite{ALICE:2017svf,ATLAS:2022iyq}), which may better isolate the underlying geometry and not be sensitive to physics effects which manifest as a bias on the produced multiplicity. Fourth, oxygen--oxygen (O+O) collisions at RHIC and the LHC~\cite{Brewer:2021kiv} would allow for the study of jet quenching in a small system, in which a single weakly-interacting nucleon in the $^{16}$O nucleus will have a much smaller impact on the overall centrality signal.

\section{Conclusion}

This paper examines a recent PHENIX measurement which found a different degree of centrality-dependent modification for $\pi^0$'s and $\gamma_\mathrm{dir}$'s in $d$+Au collisions at RHIC. The measurement strategy is ostensibly chosen to calibrate out any centrality-dependent biases with the $\gamma_\mathrm{dir}$ measurement. However, I show that the particular $\pi^0$ and $\gamma_\mathrm{dir}$ kinematic selections used in the measurement select events which arise from different distributions of Bjorken-$x$ values. As such, the PHENIX measurement is directly sensitive to the physics effects described by the color fluctuation model, in which centrality-dependent modifications (which do not arise from jet quenching) systematically increase with $x$. Using the model parameters determined from RHIC and LHC measurements reported ten years ago, without any post-diction updating, I show that a straightforward application of the model to the PHENIX kinematics gives a good description of the experimental data. After accounting for the possible role of color fluctuation effects, the evidence for any remaining final-state effects, such as from jet quenching, is significantly more limited. This finding reconciles the interpretation of the PHENIX data with that from other measurements at RHIC and LHC, which have set stringent limits on the possible amount of parton energy loss in small collision systems.

\begin{acknowledgments}
DVP acknowledges John Lajoie, Riccardo Longo, Jamie Nagle, Peter Steinberg, and Mark Strikman for useful discussions and input on an early draft of this paper. DVP's work is supported by the U.S. Department of Energy, grant DE-FG02-03ER41244. 
\end{acknowledgments}

\bibliography{apssamp}

\end{document}